\documentclass{article}
\usepackage{geometry}           
\usepackage{graphicx}
\usepackage{subfig}
\graphicspath{{./}}
\usepackage{amsmath,amssymb,amsthm}
\usepackage{bm}
\usepackage{color}
\usepackage[normalem]{ulem}

\newlength{\figwidth}
\newlength{\figwidthtwo}
\newlength{\figwidththree}
\newcommand{\fref}[1]{Fig.\,\ref{#1}}

\newcommand{\eref}[1]{Eq.\,(\ref{#1})}

\newcommand{\sref}[1]{Sec.\!~\ref{#1}}

\newcommand{\cref}[1]{Ref.\,\cite{#1}}
\newcommand{\crefs}[1]{Refs.\,\cite{#1}}

\newcommand{\ie}{{\it i.e.}\!\, }
\newcommand{\eg}{{\it e.g.}\!\, }
\newcommand{\etc}{{\it etc.}\! }
\newcommand{\etal}{{\it et al.} }
\newcommand{\apriori}{{\it a priori} }

\newcommand{\insitu}{{\it in situ} }
\newcommand{\adhoc}{{\it ad hoc} }

\newcommand{\bs}{\mathsf{b}}

\newcommand{\xs}{\mathsf{x}}
\newcommand{\ys}{\mathsf{y}}

\newcommand{\Ws}{\mathsf{W}}
\newcommand{\Xs}{\mathsf{X}}
\newcommand{\Ks}{\mathsf{K}}
\newcommand{\Ls}{\mathsf{L}}

\newcommand{\Ys}{\mathsf{Y}}
\newcommand{\Zs}{\mathsf{Z}}

\newcommand{\Uc}{\mathcal{U}}

\newcommand{\Cbb}{\mathbb{C}}

\newcommand{\epsilonb}{\boldsymbol{\epsilon}}
\newcommand{\varepsilonb}{\boldsymbol{\varepsilon}}
\newcommand{\sigmab}{\boldsymbol{\sigma}}

\newcommand{\eb}{\mathbf{e}}

\newcommand{\xb}{\mathbf{x}}

\newcommand{\pb}{\mathbf{p}}

\newcommand{\Gb}{\mathbf{G}}

\newcommand{\Fb}{\mathbf{F}}

\newcommand{\Eb}{\mathbf{E}}

\newcommand{\Pb}{\mathbf{P}}
\newcommand{\Sb}{\mathbf{S}}
\newcommand{\Ib}{\mathbf{I}}
\newcommand{\Rb}{\mathbf{R}}

\usepackage{tipa}
\newcommand{\pth}{\text{\sffamily\itshape\textthorn}}
\newcommand{\sth}{\text{\textthorn}}
\newcommand{\strain}{\boldsymbol{\epsilon}}
\newcommand{\stress}{\boldsymbol{\sigma}}

\title{\bf Predicting the mechanical response of oligocrystals with deep learning}
\author{
A.L. Frankel, R.E. Jones\footnote{rjones@sandia.gov},\, C. Alleman, J.A. Templeton\\
{\it Sandia National Laboratories, Livermore, CA 94551}
}
\date{}      

\begin{document}
\maketitle

\begin{abstract}
In this work we employ data-driven homogenization approaches to predict the particular mechanical evolution of polycrystalline aggregates with tens of individual crystals.
In these oligocrystals the differences in stress response due to microstructural variation is pronounced.
Shell-like structures produced by metal-based additive manufacturing and the like make the prediction of the behavior of oligocrystals technologically relevant.
The predictions of traditional homogenization theories based on grain volumes are not sensitive to variations in local grain neighborhoods.
Direct simulation of the local response with crystal plasticity finite element methods is more detailed, but the computations are expensive.
To represent the stress-strain response of a polycrystalline sample given its initial grain texture and morphology we have designed a novel neural network that incorporates a convolution component to observe and reduce the information in the crystal texture field and a recursive component to represent the causal nature of the history information.
This model exhibits accuracy on par with crystal plasticity simulations at minimal computational cost per prediction.
\end{abstract}

\section{Introduction}

Predicting the elasto-viscoplastic response of a particular volume of material with a limited amount of information about the microstructure and internal state is a long-standing problem \cite{matouvs2017review}.
For the case of polycrystalline metals, detailed finite element models with crystal plasticity (CP) constitutive rules have been successful in emulating the mechanical response up to failure \cite{lebensohn1993self,miehe1999computational,raabe2001micromechanical}.
Two useful concepts provide context for the results of direct simulation of materials with microstructure.
The notion of a representative volume element (RVE) was introduced by Hill \cite{hill1963elastic,drugan1996micromechanics} and is a geometrically regular sample (usually a cube) with simple boundary conditions that has all the microstructural details necessary to be representative of the material of interest and large enough that surface and region boundary effects are not pronounced. The simulation of the mechanical response of an RVE is supposed to provide \emph{effective} material properties that do not vary from sample to sample.
The statistical volume element (SVE) \cite{ostoja2006material} is a smaller sample with a characteristic feature-to-sample size ratio small enough that statistical variations exist between realization, and each instance provides \emph{apparent} material properties that converge to the effective material properties as the size of the SVE is increased.
The mean response of an ensemble of SVEs is expected to be the response of an RVE, which, in turn, is expected to be the response of the bulk material it represents.

Well-established homogenization theory exists that can provide predictions of the mechanical response of a given sample based on readily accessible microstructural information, such as grain size. 
These homogenization theories are most accurate for polycrystals with large number of grains that are not strongly textured so that the effective properties regress to the mean in the limit of small grain to sample size (separation of scales).
Theories that treat defects, inclusions, grain boundaries and other microstructure are compiled in a number of classic texts spanning decades, \eg \crefs{bakhvalov1989averaging,ostoja2007microstructural,bensoussan2011asymptotic,torquato2013random,nemat2013micromechanics,mura2013micromechanics}.
Sophisticated methods \cite{kroner1958computation,budiansky1961theoretical,hashin1963variational,castaneda1991effective,castaneda1992new} based on Eshelbian mechanics or minimum energy variational principles and simplifying assumptions evolved from these early theories and typically involve significant computation to evaluate.
Here we use the classical uniform strain and uniform stress approximations for the basis of comparison since they are widely used and relatively simple to implement.
For the elastic response of a polycrystal, Voigt \cite{voigt2014lehrbuch} and Reuss \cite{reuss1929berechnung} developed volume average theories assuming uniform strain or uniform stress states that provided complementary upper and lower bounds on the effective stress.
Since these bounds are biased in opposite ways, the Voigt-Reuss-Hill average \cite{hill1952elastic} of the two is generally expected to be a better predictor.
For the plastic response of a polycrystal, Taylor \cite{taylor1938plastic} and Sachs \cite{sachs1928plasticity} developed analogous theories to Voigt and Reuss using the uniform strain and  uniform stress assumptions, respectively.
Unlike the analytical Voigt and Reuss bounds, the Taylor and Sachs models generally require complex computations to model the viscoplastic response of polycrystals.

In this work, we employ artificial neural networks to create data-driven models of stress response that accurately incorporate the effects of microstructural variation via the initial conditions of grain orientation, shape, and topology, as encapsulated in an image of the pre-deformation microstructure.
The application of machine learning techniques to physical problems is a rapidly expanding field.
A number works are notable in using image data and convolutional neural networks to improve predictions of mechanical outcomes, and most focus on categorization or reconstruction of microstructure.
Lubbers \etal \cite{lubbers2017inferring} adapted a pre-trained convolutional neural network design from image recognition to the task of forming low-dimensional representations of microstructures to facilitate statistical reconstructions. 
Similarly, Cang \etal \cite{cang2016deep} used restricted Boltzmann machines to develop reconstructed version of microstructural images.
Kondo \etal \cite{kondo2017microstructure} also developed convolutional neural networks to regress ionic conductivity in materials directly from images of ceramic microstructures.
Chowdhury \etal \cite{chowdhury2016image} used pretrained image recognition software and other machine learning classifiers to classify the presence of dendrites in microstructure images.
Yao \etal \cite{yao2016kinetic} inferred molecular kinetic energy from density functional theory simulations of electron density in hydrocarbons using convolutional neural networks.
Hanakata \etal \cite{hanakata2018accelerated} employed a convolutional neural network model to accelerate the design of molecular graphene kirigami.
Xie and Grossman \cite{xie2018crystal} employed a convolutional network to predict the properties of crystals using a graph-based input space.
Ling \etal \cite{ling2017building} used convolutional neural networks and random forests to classify microstructure and explored generalization between data sets, number of features required, and interpretability.
Papanikolaou \etal \cite{papanikolaou2017learning} employed digital image correlation of dislocation motion, principal component analysis and machine learning techniques to classify aspects of crystal plasticity.
Liu \etal \cite{liu2018microstructural} employed a database of microstructures, a clustering algorithm, and an effective material stiffness tensor to predict plastic localization.
Beck \etal \cite{beck2018deep} developed convolutional neural networks to model sub-grid stresses in large eddy simulations of turbulent fluid flow.
Other works have considered functions with complex history-dependent behavior, as is often encountered in time-series data. 
For example, Wang \etal \cite{wangsun2018lstm} used a recurrent neural network approach to develop accurate homogenization models for coupled hydro-mechanical problems of flow in porous materials. In this study, the neural network architecture is employed for the ambitious task of reproducing the entire stress-strain history of a material given a representation of the initial microstructure.

\begin{figure}
\centering
{\includegraphics[width=0.5\textwidth]{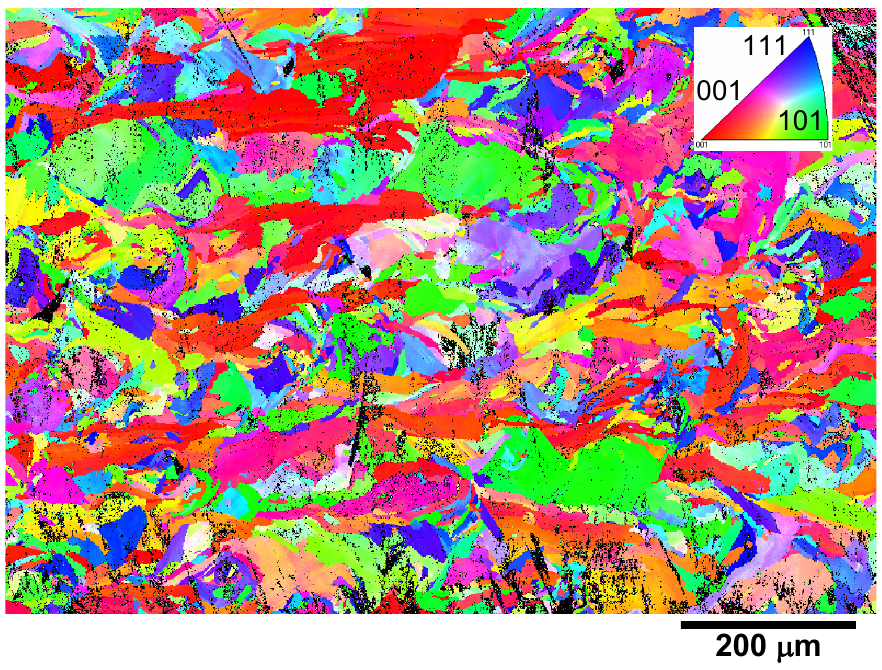}}
\caption{Electron backscatter diffractometry (EBSD) showing grain boundaries and orientations (courtesy of B. Boyce, Sandia, microstructure associated with the rapid solidification of austenitic 316L stainless steel produced by laser powder bed fusion using a ProX 200 printer from 3D Systems).
}
\label{fig:EBSD}
\end{figure}

To obtain an ensemble of responses that are statistically interesting, we focus on the mechanical response of oligocrystals (polycrystals that are not near the small grain-to-sample size limit).
In these SVE-type samples, the responses exhibit significant variations across an ensemble directly attributable to variations in grain morphology and orientation.
These materials are of technological interest  particularly with the advent of additive manufacturing which tends to create large and not well controlled grain structures with characteristic sizes that can easily approach the size of features in an as-printed part.
In the case of additive manufactured materials, the variations in behavior have multiple sources, \eg  thermal effects, defects, and porosity, but most have microstructural origins tied to the manufacturing process.
With current experimental techniques, there is a wealth of microstructural information that can be obtained prior to (and \insitu with) mechanical testing.
For example, \fref{fig:EBSD} shows the variety and complexity of grain texture and morphology data that can be obtained from techniques like electron backscatter diffraction (EBSD).
This detailed microstructural information, particularly \apriori data from non-destructive methods such as X-ray computed tomography \cite{boyce2017extreme}, can be employed to refine predictions of behavior and possible failure. This work provides a means of doing just that by providing a machine-learning based framework to predict mechanical response based on readily-obtainable microstructural information.

We briefly review homogenization approaches for polycrystalline mechanical response in \sref{sec:homogenization}.
In \sref{sec:neuralnetworks}, we describe our design of a hybrid convolutional-recurrent neural network to model the response of polycrystalline aggregates.
\sref{sec:data} describes the crystal plasticity dataset we used in this study for training and testing the neural network, and how the grain morphology and texture is made amenable for machine learning.
The results are shown in \sref{sec:results} and discussed in \sref{sec:conclusion}.

\section{Homogenization Theory for Polycrystals} \label{sec:homogenization}

For a (single) crystal, the elastic modulus tensor $\Cbb$ is defined by
\begin{equation} \label{eq:stiffness}
\Cbb = \boldsymbol{\partial}_{\strain} \stress \ ,
\end{equation}
where the stress $\stress$ and strain $\strain$ states are measured in the crystal frame with axes $\{ \Eb_i \}$. 
This stiffness tensor may be rotated to the particular orientation of the crystal in the polycrystalline aggregate $\{ \eb_i = \Rb \Eb_i \}$ via
\begin{equation} \label{eq:RC}
\tilde{\Cbb} = \Rb \boxtimes \Cbb = \sum_{i,j,k,l} \left[ \Cbb \right]_{ijkl} \Rb \Eb_i \otimes \Rb \Eb_j \otimes \Rb \Eb_k \otimes \Rb \Eb_l \ ,
\end{equation}
where $\boxtimes$ is the Kronecker product.

To obtain the apparent modulus tensor $\bar{\Cbb}$ of a polycrystal, it is not sufficient to simply volume-average the anisotropic elastic moduli for each grain.
The compatibility and equilibrium conditions between neighboring grains create non-uniform strain and stress fields that cannot be resolved analytically.
Nevertheless, two simple bounding estimates may be derived for the stiffness tensor.
The Voigt average \cite{voigt2014lehrbuch} for a polycrystal with grain volume fractions $\phi_g$, where $g$ indexes grains, is
\begin{equation}
\langle \Cbb \rangle = \sum_g \phi_g \Cbb_g
\end{equation}
and the Reuss average \cite{reuss1929berechnung} is given by
\begin{equation}
\langle \Cbb \rangle = \left( \sum_g \phi_g \Cbb_g^{-1} \right)^{-1}.
\end{equation}
The Voigt average of the stiffnesses $\Cbb$ gives an upper bound to $\bar{\Cbb}$ and corresponds to the assumption that the strain state is uniform throughout the polycrystal.
The Reuss average of the compliances $\Cbb^{-1}$ gives a lower bound and corresponds to the assumption that the stress state is uniform.
Since the grain compatibility and boundary conditions limit the accuracy of either of these assumptions, there may be a substantial discrepancy between these averages and the true elastic moduli.
As mentioned, the Voigt-Reuss-Hill approximation \cite{hill1952elastic} is simply the mean of the two bounds
\begin{equation}
\langle \Cbb \rangle = \frac{1}{2}  \sum_g \phi_g  \Cbb_g + \frac{1}{2} \left( \sum_g \phi_g \Cbb_g^{-1} \right)^{-1} 
\end{equation}
This treatment typically lowers the bias relative to the two separate estimators but is equivocal to the particular boundary conditions.

For plastic response, Taylor \cite{taylor1938plastic} and Sachs \cite{sachs1928plasticity} developed the corresponding theories to Voigt and Reuss. 
These models are generally not completely analytical and require significant computation to approximate, for example, the common problem of evolving uniaxial states of stress.
Briefly, the isostrain approximation assumes the stress $\stress_g$  in each grain $g$ is a function of a uniform deformation gradient $\Fb$, the elastic modulus tensor $\Cbb_g$, and the local material state $\Fb^e_g$ (with $\Fb \equiv \Fb^e_g \Fb^p_g$).
For a uniaxial-stress deformation, the constitutive model (such as that in \sref{sec:data}) is updated according to a prescribed $\Fb$, and a volume-averaged stress $\stress = \sum_g \phi_g \stress_g$ which is computed.
A nonlinear solver must then be employed to iterate on the value of $\Fb$ until a uniaxial $\stress$ is obtained, subject to a constraint on $\Fb$ from the displacement-control of the effective boundary condition.
In the isostress condition, a uniform stress $\stress$ is prescribed, and a volume-averaged displacement gradient $\Fb = \sum_g \phi_g \Fb_g$ is computed from the constitutive update.
The value of $\stress$ must then be iterated until the constraint on $\Fb$ due to the effective boundary condition is satisfied.
Either strategy (isostrain or isostress) must be applied step-wise over a finite sequence of time-steps to approximate the evolution of the material state for a given deformation history.
The resulting approximations are systematically biased, with the isostrain approximation of Taylor giving and effective upper bound on the flow stress, and the isostress approximation of Sachs giving the corresponding lower bound.
As in the elastic case, an approximation can be constructed based on averaging the two bounds, reducing the bias in an essentially empirical fashion.

In Section \ref{sec:results}, the neural-network approach developed here is compared to these classical treatments, with separate evaluations for the elastic and elasto-viscoplastic responses.
Specifically, it is shown that the neural-network approach naturally eliminates the biases exhibited by the other homogenization techniques without the need for \adhoc averaging.

\section{Neural Networks} \label{sec:neuralnetworks}

The basic neural network is a two-dimensional feed-forward network, often called a multilayer perceptron (MLP) \cite{rosenblatt1961principles}. 
It is a non-linear model that can scale to handle arbitrary complexity rapidly and, hence, is one of the most commonly employed NN architectures.
As with all NNs, it has an input layer and an output layer, each with a node per scalar, and an arbitrary number of intervening layers.
Adjacent layers of nodes are fully/densely connected in the sense that the state of the nodes of a layer is the vector $\xs_i$ of outputs from the previous layer multiplied by a weight matrix $\Ws$, added to a threshold/bias vector $\bs_i$ (both defined per layer, here index by $i$), then mapped through an activation function $f(\Ws_i \xs_i + \bs_i)$ applied component-wise. 
In this work we employed ramp-like activation function, the sharp C0 rectified linear unit (ReLU) \cite{resteghini2013single} and the C1 softplus \cite{dugas2001incorporating} activation functions, with similar results.
Adding layers between the input features and the output targets, models of arbitrary complexity may be developed but at the cost of requiring larger amounts of training/calibration data to determine the unknown weights $\Ws_i$ and biases $\bs_i$.
Using now standard algorithms, such as the stochastic gradient descent algorithm \cite{robbins1985stochastic}, the unknown parameters can be optimized to maximize the accuracy of the NN model on held-out test data.

Although MLPs are quite effective in response prediction for a modest number of inputs, it is usually impractical to train an MLP to perform regression or classification on image data, such as a voxelated realization of a polycrystal in the application at hand (refer to \fref{fig:data}a).
The number of parameters required to feed each voxel into an MLP of sufficient complexity would currently require prohibitively large amounts of data and computing power.
Furthermore, many features of interest in a given image have local, spatial correlations, and it typically effective to process subsets of voxels simultaneously since distant regions of an image are usually uncorrelated with each other.

In this section we briefly describe the proposed hybrid neural network, \fref{fig:architecture}, and its sub-networks and their constituent layers.
The proposed network is composed of: (a) an image processing component (a {\it convolutional} network, yellow and green)to assimilate the information in the initial microstructure (``image'' input, red) relevant to the stress response, and (b) a neural network with feedback (a {\it recurrent} network, blue) to emulate the history dependence of the stress (output, orange) on the strain (``history'' input, red).
In this section we give an overview of the hybrid architecture in the context of the more traditional applications of its components. 
The particular component parameters, such as specific kernel size and pooling type, will be given in the Results section.

For a more detailed description of the various and now standard components of our NN refer to the texts in \crefs{nielsen2015neural,Goodfellow-et-al-2016}, reviews \crefs{egmont2002image,lecun2015deep}, and the documentation for Keras and Tensorflow libraries used in this work \crefs{keras,tensorflow}

\subsection{Convolutional neural network}

To resolve the issues that make application of MLPs to image data impractical, the convolutional neural network (CNN) \cite{lecun1995convolutional} architecture was developed.
Convolutional neural networks are typically used in handwriting translation, face recognition and other image processing tasks such as reconstruction.
A simplified schematic of a CNN is shown in \fref{fig:architecture} in yellow.
The image data can be represented as a matrix $\Xs$ of dimensions $n + m$, where $n$ is the number of spatial dimensions and $m$ are the number of image components, which we take to be the components of the crystallographic orientation vector $\pth(\xb_I)$, described in detail in \sref{sec:texture}, at each voxel $I$.
As in standard image and signal processing, a discrete convolutional kernel $\Ks$ is applied to the image, resulting in a new image $\Ys = \Ks * \Xs$ where $*$ is the usual convolution operator.
The kernel is compact in each spatial dimension and acts on rectangular subsets of pixels, effectively eliciting local features of the image.
Similar to an MLP, an activation function $f$ is then applied to each pixel of the output image to produce $\Zs = f(\Ys)$.
As in an MLP, multiple convolutional layers may be applied in succession, but it is also often effective to introduce a pooling operation, in which the image is reduced in size by taking a norm of neighboring sets of pixels and returning the filtered result into a new image.
For example, the max-pooling operation applied over a 2$\times$2 window in a two dimensional 10$\times$10 image will find the highest value of the pixel in each disjoint 2$\times$2 set and return each maximum in a new 5$\times$5 image.
This operation forces the network to compress the information in the image to a smaller set of features.
The weight-sharing of the convolution operations and the image reductions thus provide an efficient way of processing image data in a neural network while requiring much smaller set of meta-parameters than a direct application of a MLP.
Multiple replica filters with independent kernel components can be applied to input image to increase the richness of the derived features.

\subsection{Encoder}

The output of the CNN, which is structured 3 dimensional image data in this case, may be flattened to a vector to be compatible with the input of MLP which, in this design, is a encoder.
The encoder is a data compression unit usually part of a self-consistent autoencoder used to generate artificial realizations similar to the training data.
The encoder, colored green in \fref{fig:architecture}, is simply a MLP that decreases in width (number of nodes layer to layer) in order to compress the information content of the data.

\subsection{Recurrent neural network}

\begin{figure}
\centering
\includegraphics[width=0.65\textwidth]{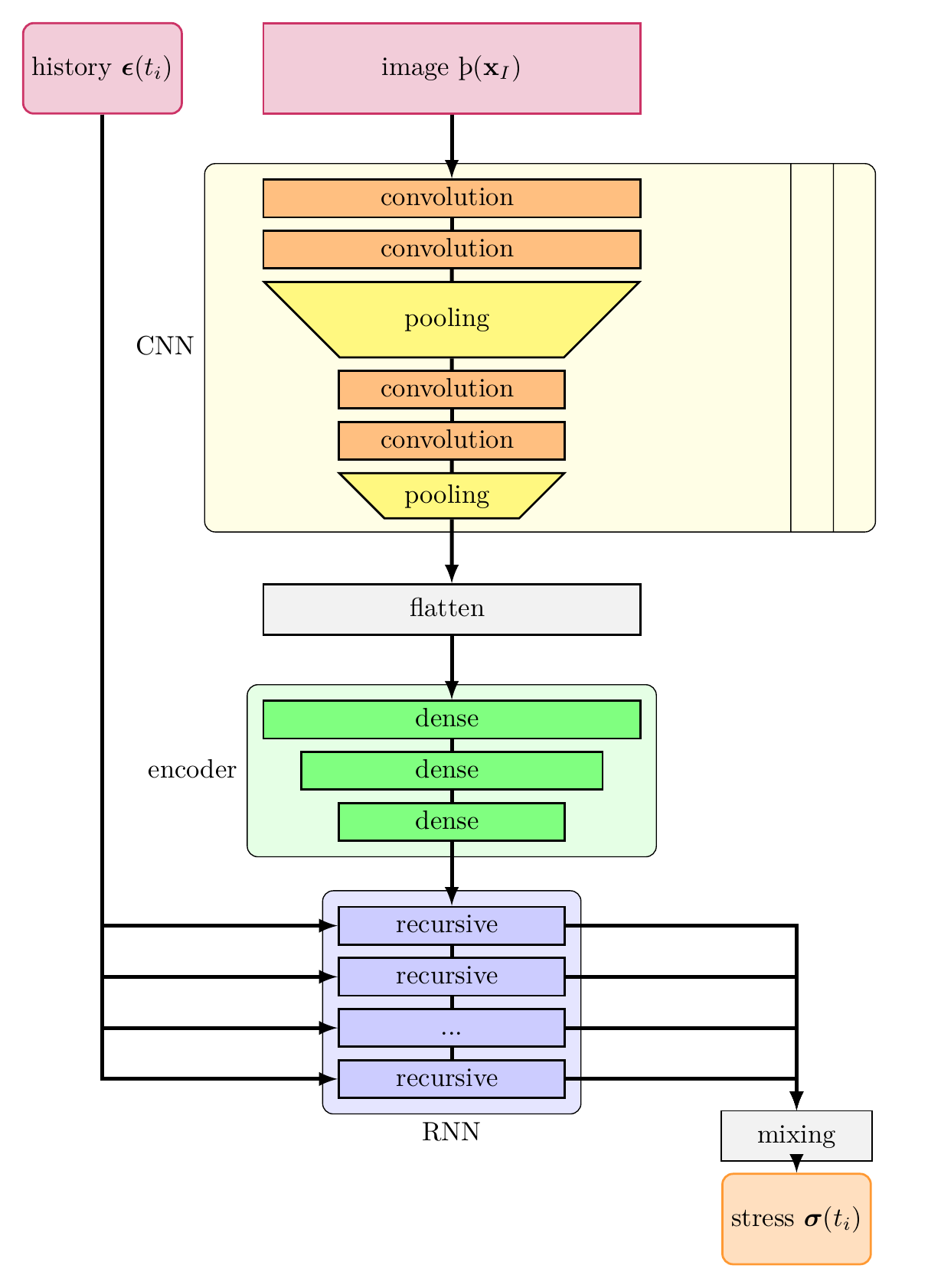}
\caption{Hybrid neural network architecture with convolutional neural network (CNN, yellow), encoder (decreasing width feed forward NN, green), and recurrent neural network (RNN, blue, shown in an ``unrolled'' diagram)  components.
Note that independent CNNs are simultaneously applied to the image and their output is combined in the flatten operation that takes the spatially correlated outputs and produces a vector of this on-grid data.
The inputs are the strain history $\epsilonb_i = \epsilonb(t_i)$ over a sequence of times $t_i, i=0,n$ and initial microstructure $\pth_I = \pth(\xb_I)$ at the image voxels $\xb_I$ (red). 
The output is stress $\sigmab_i$ at corresponding times (orange).
}
\label{fig:architecture}
\end{figure}

The output of the encoder of image data is then fed to a recurrent neural network (RNN) colored blue in \fref{fig:architecture}.
The general class of RNNs are particularly suited to modeling history dependent data since its feedback structure accounts for causality.
While a CNN takes in an image of locally structured data, a RNN takes in a sequence of data, in this case the time series of applied strains $\epsilonb_i, i=1,n$ and resulting stress $\sigmab_i$.
An individual recurrent layer takes the state of the previous time step $\ys_i$ and, along with a weighted version of the contemporary input $\epsilonb_i$, to predict the output in the subsequent time step: $ \ys_{i+1} = f(\ys_i + \Ws \epsilon_i)$.
Similar to how a CNN processes spatial data, the same weights $\Ws$ and activation $f$ are used with each time step, thus reducing the number of parameters necessary to process the information in the series data.
This design avoids the high-dimensionality of attempting to predict the entire time series simultaneously by taking advantage of the time correlation in the data.
The weights $\Ws$ and state variables $\ys_i$ are trained by comparing the prediction error $ || \sigmab_i - \ys_i || $ sequentially at each time step $i$ and updating the parameters to reduce the observed error. 
An RNN encodes time dependence through evolving state variables $\ys_i$ and thus avoids the need for observing the full history at once to make predictions.
In this work we employ a more complex but widely adopted version of an RNN composed of long short-term memory units (LSTM) \cite{hochreiter1997long}.
The LSTM includes a set of additional internal variables and weights that increase the influence of data from earlier in the series on the current state of the network.
Similar to the CNN architecture, the output from a recurrent neural network (RNN) is fed into an (linear, mixing) MLP to increase the feature richness before yielding the output.

\subsection{Illustrative example} \label{sec:toy}
To explore and illustrate the proposed architecture without limitations on sample size and image resolution inherent in the costly crystal plasticity simulations, we constructed a simple analog in which we can compare the performance the proposed neural network against a model where the subset of image information influencing the output (the so-called feature manifold) is known. 
We created synthetic two-dimensional microstructures by randomly coloring a Delaunay triangulation of a unit square from a random point cloud drawn from $\Uc(0,1)^2$. 
The density of points was such that the realizations had  between 10 and 28 grains.
Each triangular grain was assigned a scalar orientation $\sth \in \Uc(0,1)$ and hence each pixel in a structured $n\times n$ background image grid had a orientation $\sth_I$ based on the location of the pixel center $\xb_I$.
From this discretized field we created ``hidden'' features such as the average orientation $h =  \frac{1}{N} \sum_{I} \sth_I$, where $N=n^2$ is the number of pixels.
Inspired by the general work of Kalidindi \cite{kalidindi2015hierarchical} on characterizing microstructure, we generated hidden features from the average, minimum, maximum and standard deviation of the per pixel orientation $\sth_I$, the per grain volume fraction, and the per pixel grid Laplacian of orientation $\Ls * \sth$ where $\Ls$ is the usual second-order finite-difference kernel.
We also examined the replica averaged on-grid 2-point correlation of $\sth_I$, which results in a vector which we normalize by the self-correlation.
The first element of this vector is the correlation of the orientation of a pixel with that of pixels 1 pixel away, the second element is the correlation for pixels 2 pixels away, \etc\, and the  mean of this vector is an effective correlation length.
We connected these inputs to a stress-like output $\sigma$ via a linear relation
\begin{equation} \label{eq:toy_model}
\sigma = E \big( 1+ \sum_a \alpha_a h_a \bigr) \epsilon \ ,
\end{equation}
where $h_a$ is the $a$-th hidden feature, $E$ and $\alpha_a$ are parameters, and $\epsilon$ is a strain-like input.
With this simplified problem we can test the hybrid NN components independently since we know all the ``hidden'' features $\{ h_a \}$.

The proposed architecture, illustrated schematically in \fref{fig:architecture}, has a large number of meta-parameters, such as the number of convolutional layers, the width of the kernel per layer, \etc\
In fact there are too many for a comprehensive parameter study.
At first we tried to find optimal architecture via integer optimization of the meta-parameters via the SOGA algorithm \cite{eddy2001effective}.
This genetic optimization of network parameters produced good networks but no perceptible trends over the space of possible networks.
So we employed limited parameter studies, such as changing the number of convolutional layers (for fixed kernel and pool widths) and the number of encoder layers (resulting in a fixed dimension output) in the image processing component.
This particular study showed that accuracy significantly improved up to 3 convolutional layers  but showed little sensitivity to number of encoder layers.
Since there seems to be considerable redundancy in the hybrid NN architecture, we also explored using L1-regularization to effectively deactivate neural nodes with small weights/low participation in the output.
However, using regularization penalties in the range  10$^{-6}$--10$^{-3}$ (at the upper end the L1 penalty dominates) did not create more accurate networks.
In addition, for three image resolutions, 10$\times$10, 20$\times$20, 40$\times$40, that resolved the grain boundaries and also lead to feasible calculations, we found little sensitivity in optimal kernel and pool widths to image size.
Smaller images under-resolve the grain boundaries and limit the number of possible pooling layers.
For the remaining studies we used 20$\times$20 images of the initial microstructure.

\begin{figure}
\centering
{\includegraphics[width=0.55\textwidth]{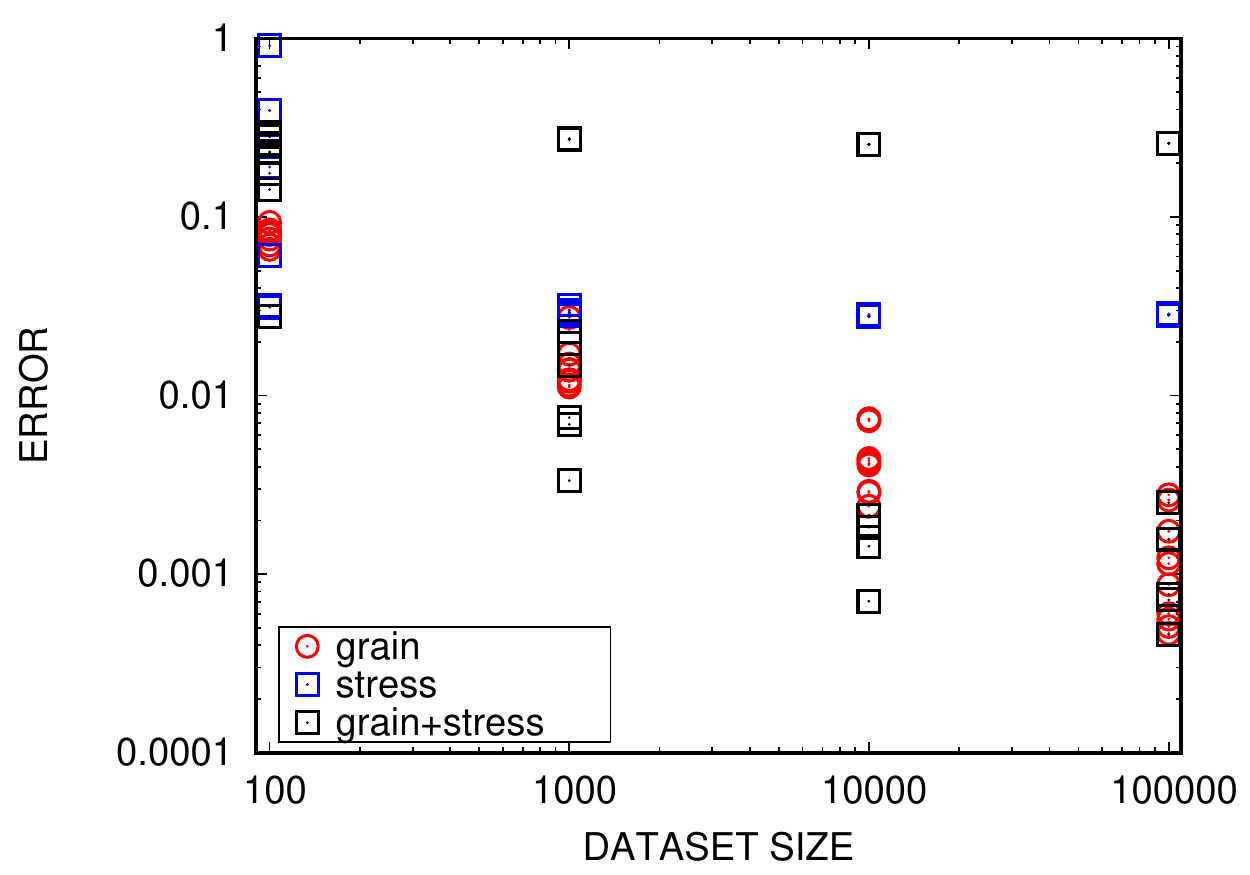}}
\caption{Convergence of the predictions of a NN trained on stress-strain data to a hybrid NN trained on stress-strain data and initial microstructure and the separate image (grain) and history (stress)  components with data set size}
\label{fig:toy_data_size}
\end{figure}

Using a baseline network with: 16 independent filters of 3 convolutional layers ( kernel width: 4 pixels, max pooling width: 2 pixels) and 3 encoder layers reducing to 1 assumed feature transmitted to the RNN, \fref{fig:toy_data_size} shows convergence with increasing data set size.
Ten replicas of this NN with randomized initial weights and thresholds were trained which resulted in trends and a certain amount of scatter in the test errors for the image component NN (yellow and green) trained on the actual hidden information (grain), the history component NN (blue) trained without image information (stress), and the complete NN all illustrated in \fref{fig:architecture}.
We used an 80/10/10 split of the image and stress-strain data into training, testing and cross-validation datasets. 
It is apparent from \fref{fig:toy_data_size} that by 1000 samples the history component alone cannot reduce the error in the predicted stress, while the image component (grain) alone, trained on the hidden feature, is still learning up to the largest data set (100,000 images) we explored.
Overall network convergence can be stymied by limitations of one of the components, here is seems that learning the hidden image features is more complex than learning the history trend.
By 1000 samples the hybrid NN is out-performing the history component trained on the same stress-strain data.
Beyond 1000 samples, the hybrid architecture is at least an order of magnitude more accurate, but there is evidence of stalling due of the well-known vanishing gradient problem where most/all the activations are in a zero state \cite{hochreiter2001gradient} for some of the trained NNs. 
For this study we used the average orientation as the sole feature, but these results are typical of any single feature of the ones we constructed.

\begin{figure}
\centering
{\includegraphics[width=0.55\textwidth]{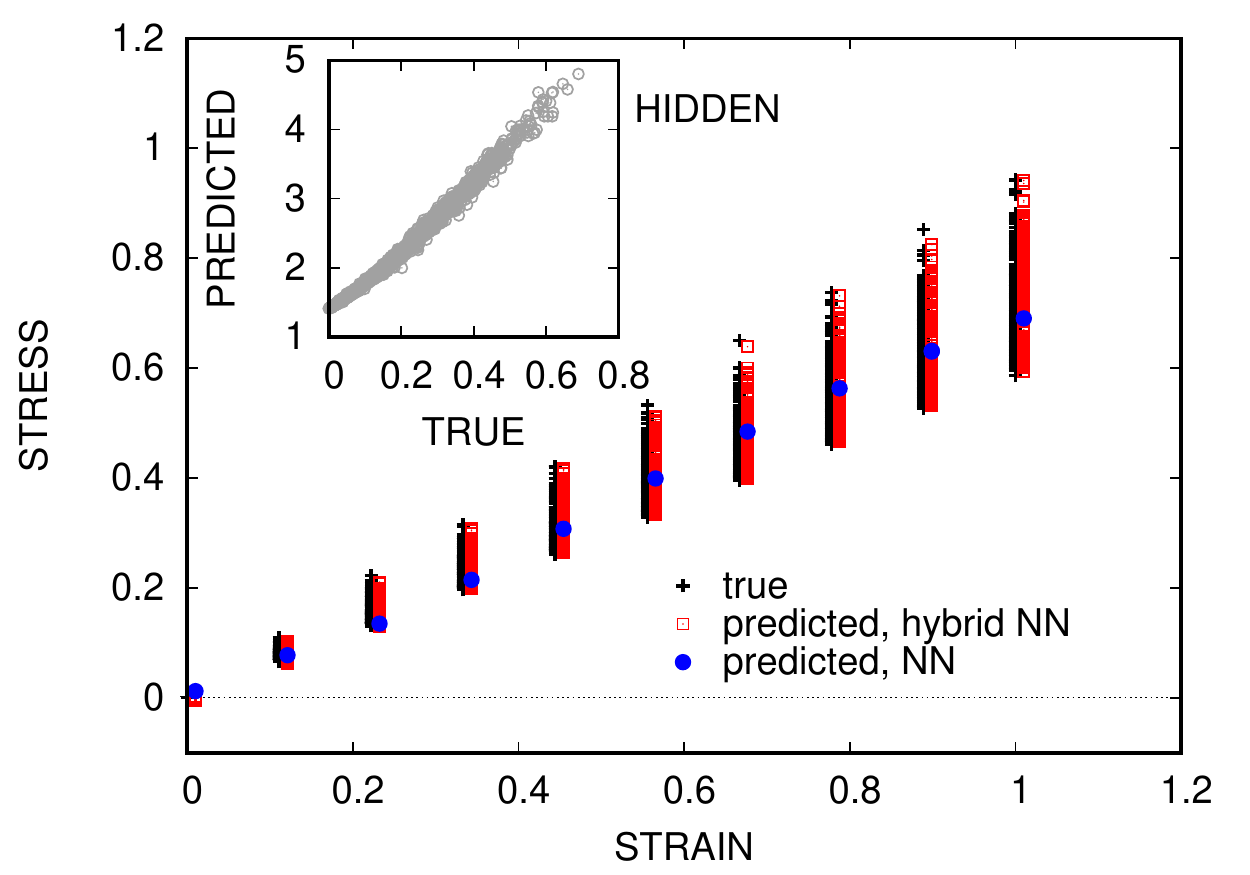}}
\caption{Comparison of the predictions of a NN trained on stress-strain data to a hybrid NN trained on stress-strain data and initial microstructure.
The inset shows the correlation of the learned ``hidden'' and the true ``hidden'' feature $h$.
}
\label{fig:toy_prediction}
\end{figure}

\fref{fig:toy_prediction} illustrates how effectively the hybrid NN incorporates the image data.
For this example we chose $h_1$ to be the average orientation for each microstructure and $\alpha_1 = 0.1$.
Given the form of \eref{eq:toy_model} we see the true response fans out in a linear fashion from the zero stress-zero strain point.
The hybrid NN captures this variation with a test error 0.002 (for dataset size  10,000), whereas the history component alone (without the image information) can only represent a mean trend with test error 0.028.
This is due to the fact that the trained, assumed hidden feature is highly correlated with the true hidden feature, but not identical to it, as can be seen in the inset of \fref{fig:toy_prediction}.
Again, the performance of the hybrid NN was similar in this test of one true hidden and one assumed hidden feature across all the hidden features we generated.

\begin{figure}
\centering
{\includegraphics[width=0.55\textwidth]{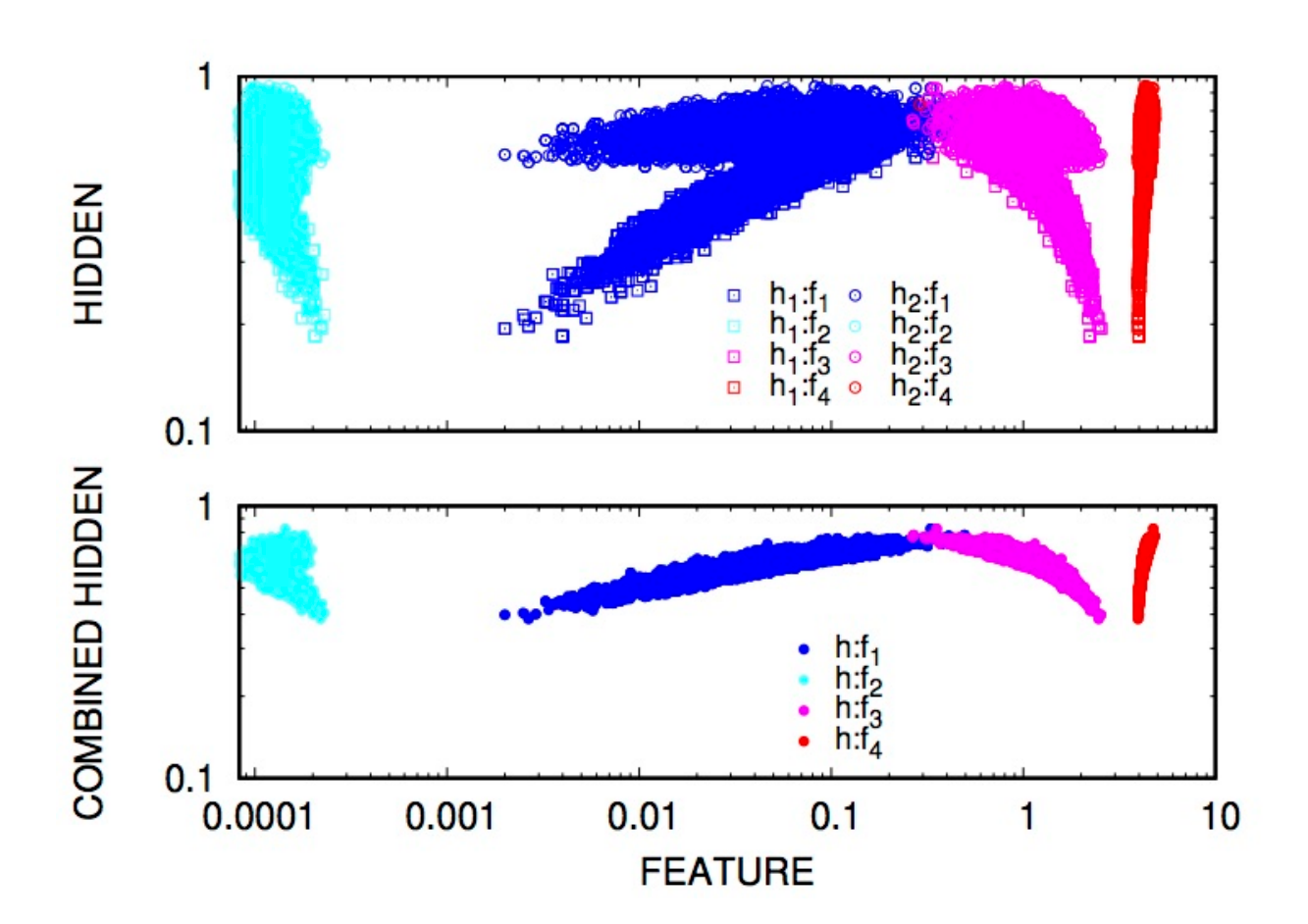}}
\caption{Correlation of hidden ($h_1$: average orientation, and $h_2$: maximum Laplacian of orientation) and 4 assumed features.
Correlations of individual hidden features (upper panel) with the assumed features, $h_1$: 0.83, -0.09, -0.93, 0.90, and $h_2$: 0.29, -0.18, -0.39, 0.41.
Correlation of the combined hidden feature $\alpha_1 h_1 + \alpha_2 h_2$ (lower panel): 0.82, -0.16, -0.95, 0.94.
}
\label{fig:toy_hidden_correlation}
\end{figure}

\fref{fig:toy_prediction} illustrates the performance of the network for a single hidden and single assumed image feature; we also explored how the hybrid NN behaves with a dependence on multiple image features and where the number of hidden and assumed features are mismatched.
We explored dependence on four uncorrelated features: (1) the average orientation, (2) the maximum grid Laplacian, and (3) the standard deviation of the orientation.
(The minimum correlation of the set of features was between the average orientation and the standard deviation: $R$=0.01, and the maximum was the maximum grid Laplacian with the standard deviation: $R$= 0.38.)
Generally speaking we found that using between 1 and 4 assumed features we could achieve test error on the order of 0.01 or lower, and typically had at least one assumed feature highly correlated ($R >$ 0.9) with each of the outputs but not always.
For simpler to visualize case of 2 hidden features and 4 assumed features,
\fref{fig:toy_hidden_correlation} shows correlation of the individual hidden features and the assumed ones.
Here the best correlation for the first hidden feature is -0.93 but only 0.41 for the second.
Examining the form of the model, \eref{eq:toy_model}, explains how good accuracy is achieved with imperfect correlation with the individual hidden features.
The lower panel of \fref{fig:toy_hidden_correlation} shows that at least two of the assumed features are correlated $R > $ 0.9 with the combined feature $h =\sum_a \alpha_a h_a$.
This result implies that ascertaining the effective dimensionality of the unknown feature manifold using the rank of trained assumed features may be possible.

\section{Polycrystalline Dataset} \label{sec:data}
In this section we describe the set of microstructural realizations used to generate stress data and the representation of the texture in each realization.
Subject to common simplifying assumptions, the material model is representative of an annealed austenitic stainless steel.

\subsection{Polycrystalline Realizations} \label{sec:realizations}
\begin{figure}
\centering
\includegraphics[height=0.4\textwidth]{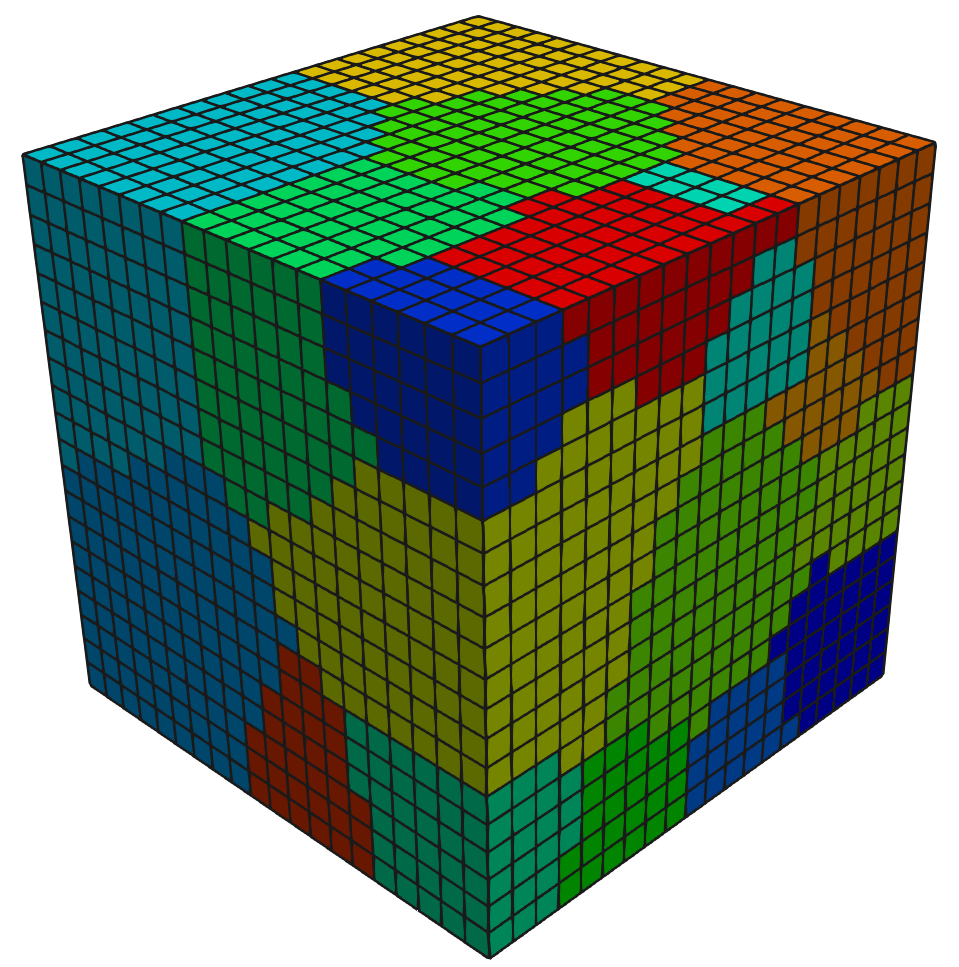}
\includegraphics[height=0.4\textwidth]{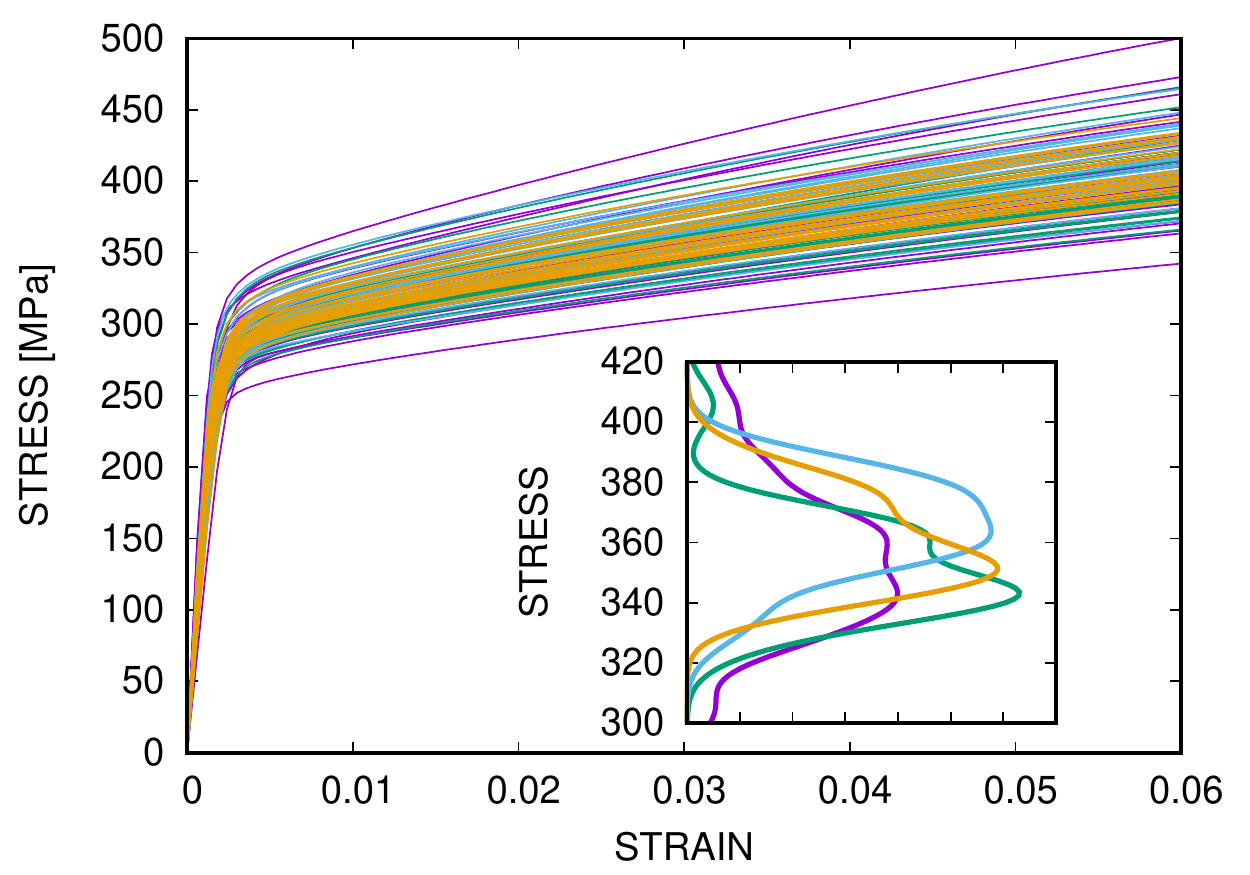}
\caption{Polycrystalline realization and stress-strain data.
The grain realization is colored by the (100) component of the Rodriguez orientation vector.
The stress-strain plots are colored by grain realization and the inset shows the distribution of stress of each realization of the morphology over all the texture samples at 3\
}
\label{fig:data}
\end{figure}

For this is study we examine the response of a 1 $\mu$m$^3$ cube of polycrystalline steel with a single face-centered cubic (FCC) phase.
To create each polycrystalline realization we partitioned the cube into grains to emulate a realistic polycrystalline morphology using the DREAM.3D software package \cite{Groeber2014}, and then assigned orientations to each crystal to create textures.
For computational efficiency, we used a 20$^3$ element structured mesh  with voxelated grain boundaries, \fref{fig:data}a shows a realization.
Thirty-one individual oligocrystals were constructed to have approximately log-normal distributions of grain size.
The number of grains per realization in this dataset ranged from 7 to 24.
For each grain morphology, 30 textures were sampled uniformly over the special orthogonal group SO(3) for each grain. For 10 of the polycrystals, an additional 570 textures were sampled.

The response of each crystal followed an elastic-viscoplastic constitutive relation based on well-known meso-scale models of single crystal deformation  \cite{taylor1934mechanism,kroner1961plastic,bishop1951xlvi,bishop1951cxxviii,mandel1965generalisation,dawson2000computational,roters2010overview} using parameters representative of steel.
For the crystal elasticity, we employed a linear relation between the second Piola-Kirchhoff stress $\Sb$ in the intermediate configuration and the elastic Lagrangian strain $\Eb^e$,
\begin{equation}
\Sb = \Cbb : \Eb^e
\end{equation}
Cubic symmetry is assumed, and the independent components of the elastic modulus tensor $\Cbb$ are $C_{11}, C_{12}, C_{44}$ = 204.6, 137.7, 126.2 GPa.
In each crystal, plastic flow can occur on any of the 12 FCC slip planes, and we employed a common power-law form for the slip rate relation
\begin{equation}
  \dot{\gamma}_{\alpha}=\dot{\gamma}_0\left|\frac{\tau_{\alpha}}{g_{\alpha}}\right|^{m-1}\tau_{\alpha} \ ,
\end{equation}
driven by the shear stress $\tau_\alpha$ resolved on slip system $\alpha$.
The reference slip rate was chosen to be $\dot{\gamma}_0$ = 1.0 s$^{-1}$, the rate sensitivity exponent was $m = 20$, and the slip resistance $g_{\alpha}$ was given the initial value $g_{\alpha}$ = 122.0 MPa. 
The slip resistance evolved according to \cite{Kocks1976, mecking1976hardening}
\begin{equation}
\dot{g}^\alpha = (H-R_d g^\alpha) \sum_\alpha |\dot{\gamma}^\alpha|
\end{equation}
where the hardening modulus was chosen to be $H = 355.0$ MPa and the recovery constant was $R_d = 2.9$. 
See \cref{jones2019machine} for more details.

Each oligocrystal realization was subjected to quasi-static uniaxial tension at a constant engineering strain-rate of $\dot{\epsilon} = 1$ s$^{-1}$ up to 6\
The volume-averaged stress was computed from each realization, and the apparent Young's modulus was approximated by dividing the change in stress by the initial strain increment of 0.0006.
Representative stress-strain data is shown in \fref{fig:data}b.
The mean elastic modulus is 177.6 GPa and has a coefficient of variation of 11.8\

\subsection{Texture Representation} \label{sec:texture}
The crystal orientation for each grain in every oligocrystal  was sampled uniformly on SO(3), which may be represented as an orientation matrix $\Rb$, as in \eref{eq:RC}, with the Euler-Rodriguez formula:
\begin{equation}
\Rb(\theta\pb) = \exp(\theta \Pb) = \Ib + (\sin \theta) \Pb + (1-\cos \theta) \Pb^2
\end{equation}
where $\theta$ is the rotation angle, $\pb$ is the (unit) axis vector, and $\Pb = \varepsilonb \pb$ with $\varepsilonb$ being the 3rd order permutation tensor.
Given the cubic symmetry in the crystal, there are 24 equivalent representations of the same texture.
Denoting $\Gb$ as an element of the cubic symmetry group, defined by its generators $\{ \Rb(\pi/2 \Eb_1), \Rb(2 \pi/3 (\Eb_1+\Eb_2+\Eb_3)), \Ib-2\Eb_2\otimes\Eb_2, -\Ib \}$, we have
\begin{equation}
\Rb' = \Rb \Gb
\end{equation}
where $\Rb' \boxtimes \Cbb$ has the same components as $\Rb \boxtimes \Cbb$.

Since we aim to develop an efficient approximation of the underlying physics from the textured realizations directly from a relatively small amount of data, it is important to find a compact common basis for the texture representation.
In this work we choose to use the axis-angle representation of a texture
\begin{equation}
\pth = \theta \pb
\end{equation}
where $\theta$ is the rotation angle and $\pb$ is the axis of rotation for the texture.
This is a convenient embedding of the 3 parameters necessary to uniquely specify a texture.
In order to reduce the space of admissible textures, each of the 24 cubic symmetry operations is applied to the orientation matrix, and the axis-angle representation of each texture is computed.
The axis-angle pair that is selected is the one that lies in the positive octant (\ie each component of $\pth$ is greater than zero) and has the smallest rotation angle.
This folding of the orientation vector $\pth$ increases data density and hence is a form of data augmentation.
This procedure applied to voxelated realizations  result in a 20$\times$20$\times$20$\times$3 image inputs to the convolutional neural network (CNN) shown in \fref{fig:architecture}.

\section{Results} \label{sec:results}
In this section we show the results of training a neural network architecture to make predictions of the crystal plasticity stress-strain curve directly from the initial polycrystalline microstructure.
The network is shown in \fref{fig:architecture} and operates on 20$\times$20$\times$20 three dimensional (3D) images of the microstructure (colored by the 3 components of the orientation vector field) and the strain history.
The design was guided by the investigation in \sref{sec:toy}.
In this case, the (yellow) CNN is composed of:
a 3D convolutional layer with kernel size 1, a 3D convolutional layer with kernel size 2, a max pooling layer with pool size 4, two more 3D convolutional layers with kernel size 2, and finally an average pooling layer with pool size 2.
We used 16 replica filters of this type.
The encoder (green) was a single layer taking the 16$\times$10$^3$ image outputs to 10 nodes.
The (blue) RNN consisted of as many LSTM layers as time inputs (1 step was used in the elastic predictions and 10 in the plastic predictions).
The output of the overall network culminated in a mixing layer that reduced the 10 outputs of the RNN to the single stress output per time step.
The number of layers of each convolution stage, kernel size, and number of filters were tuned manually and sequentially based on which architecture appeared to give the best improvement in cross-validation error until no further discernible improvement could be made.
For instance, the first convolution layer in each network has kernel size 1, essentially acting as a purely local operation.
This has the effect of expanding the input orientation vector into a richer feature space that were then convolved spatially to incorporate interactions with neighboring voxels.

\subsection{Predicting the elastic modulus}
\begin{figure}
\centering
\includegraphics[width=0.68\textwidth]{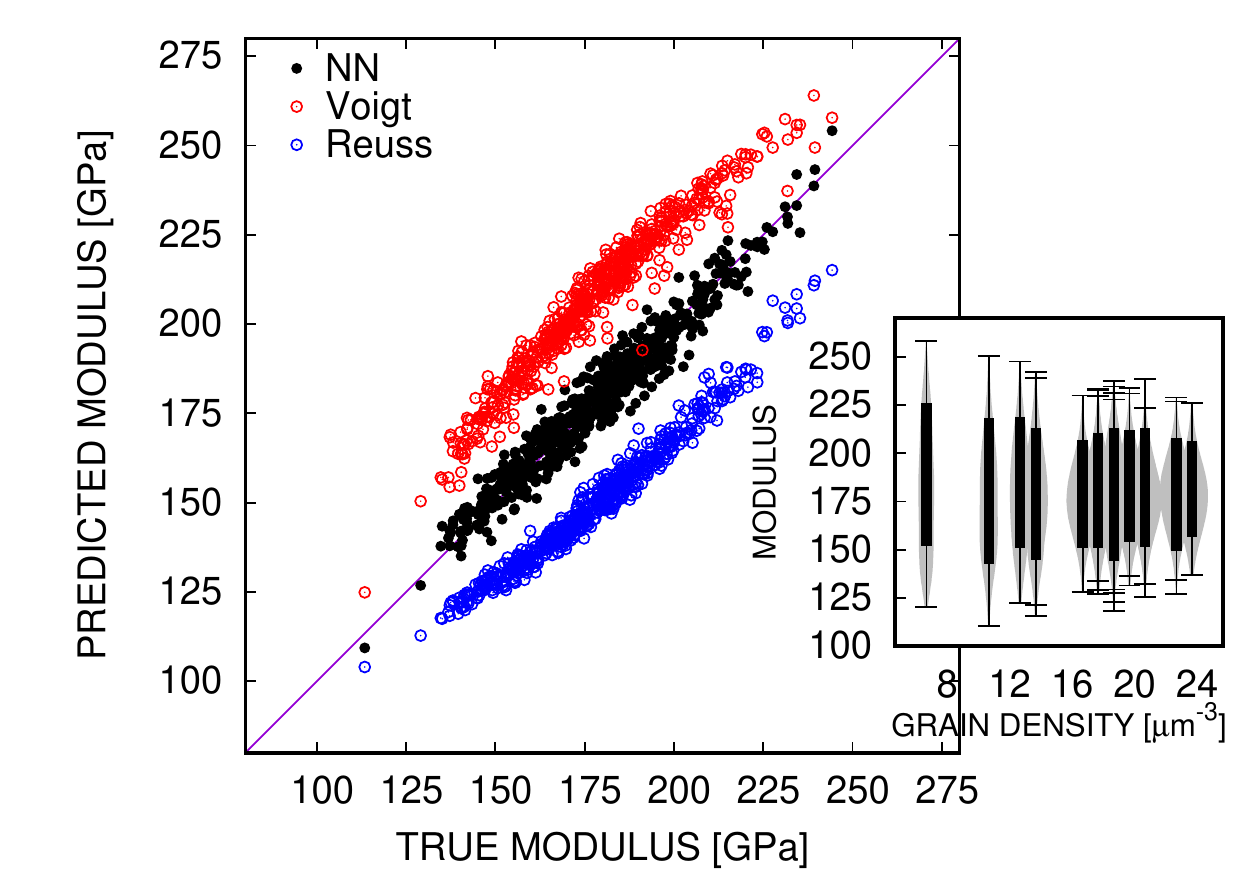}
\caption{
Variance of the tensile modulus with respect to grain topology and textures.
Prediction of the tensile modulus.
The correlation between the predictions and the test data is 0.97.
}
\label{fig:E}
\end{figure}

To test the predictive capabilities of the proposed data-driven homogenization model, we first tried to predict the first step in the deformation path which is effectively the same as predicting the elastic modulus.
The network was trained with cross-validation, holding out an individual grain topology and evaluating the network on the removed textured samples.
This leave-one-out validation was repeated for 20 different grain topologies to test the generalizability of the network predictions, leading to 20 different trained networks and 30 samples per network of an unseen grain topology for testing per network.
\fref{fig:E} shows the comparison of the neural network predicted moduli against the Voigt and Reuss averaged predictions.
The neural network predictions are highly correlated with the observed stiffness predictions, with a correlation coefficient of $R=0.974$ and show far less bias.
For comparison we also calculated the correlation coefficient $R=0.988$ for the Hill average.
We also calculated the bias in each of the predictions.
The Voigt model overestimates by 16.6\
The Hill bias is  -0.490\
The fact that the best empirical estimate and the data driven one are on par is encouraging. 
It is plausible that given different boundary conditions the Hill average of a constant stress and a constant strain predictor will fail to give as high a correlation, while the machine learned technique should be insensitive to the boundary conditions as long as there are representative cases in the training data.

We also used the trained neural networks for forward uncertainty propagation to observe the effects of grain structure and texture on the distribution of elastic moduli.
One thousand different random textures were assigned to each held out grain topology, and the neural network was used to predict the elastic modulus for each polycrystal.
The inset of \fref{fig:E} shows the spread in possible outcomes for the different topologies as a function of the number density of grains in the structure.
It is clear that the mean of the elastic modulus does not vary significantly between the structures, and the distribution retains its relative shape, but the overall variance in the observed moduli tends to decrease with an increasing number of grains.
This result suggests that the largest driver for variability in the response of the oligocrystals we examined and the neural network predictions is the variety of textures.
This is consistent with the expectation that the homogeneous limit of an arbitrarily large number of grains would show no significant variability in mechanical properties.

\subsection{Predicting the onset of plastic flow}

The variability in the simulated stress-strain curve is due entirely to variations in the microstructure; thus the microstructural features derived from a CNN may also be used to predict the entire stress-strain evolution.
To that end, we feed the output from a CNN and a vector of strain values to the input of an RNN to predict the stress-strain response, as illustrated in \fref{fig:architecture}.
We focus on the initial part of the deformation, where the oligocrystals yield and begin to undergo plastic flow.

\begin{figure}
\centering
\includegraphics[width=0.55\textwidth]{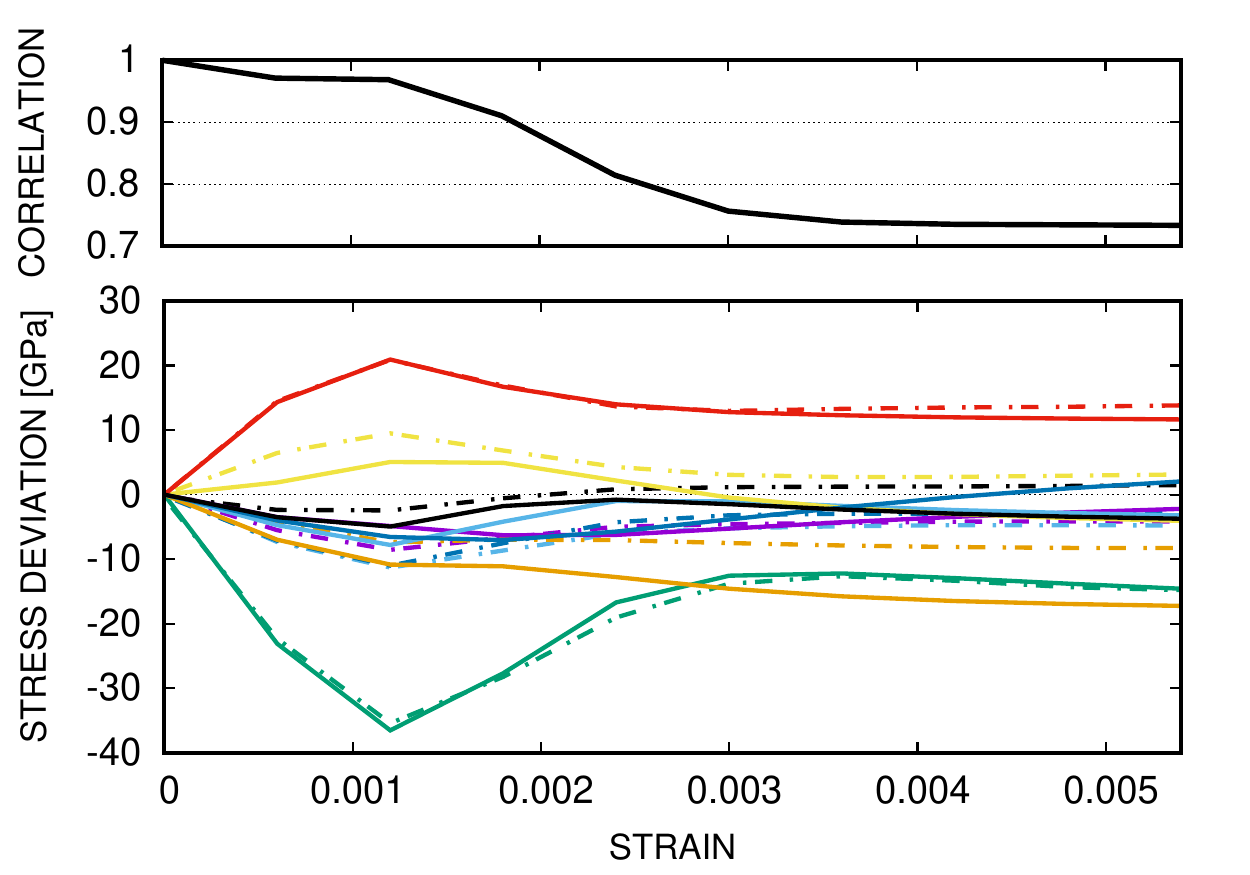}
\caption{Prediction of the stress evolution: deviation from the mean response of the ensemble.
Solid lines: simulation data, dashed lines: neural network predictions.
}
\label{fig:RNN_realizations}
\end{figure}

In this case, we augment the  dataset by adding rotations of the microstructure using symmetry operations, \eg a reflection through the tensile axis.
Five of the microstructures (and their rotations) were held out and used as testing data.
\fref{fig:RNN_realizations} shows that the predictions of this model are well correlation with the data but accuracy decreases with time due to an accumulation of errors.
We also observe that there is better accuracy for realizations with larger deviations from the mean trend.
Correlation overall decreases if more of the strain history is used due to the increasing emphasis on the fully plastic regime (not shown for brevity). 
This can be offset by emphasizing the elastic-plastic transition  by increasing the data density in this critical regime.

\begin{figure}
\centering
\includegraphics[width=0.65\textwidth]{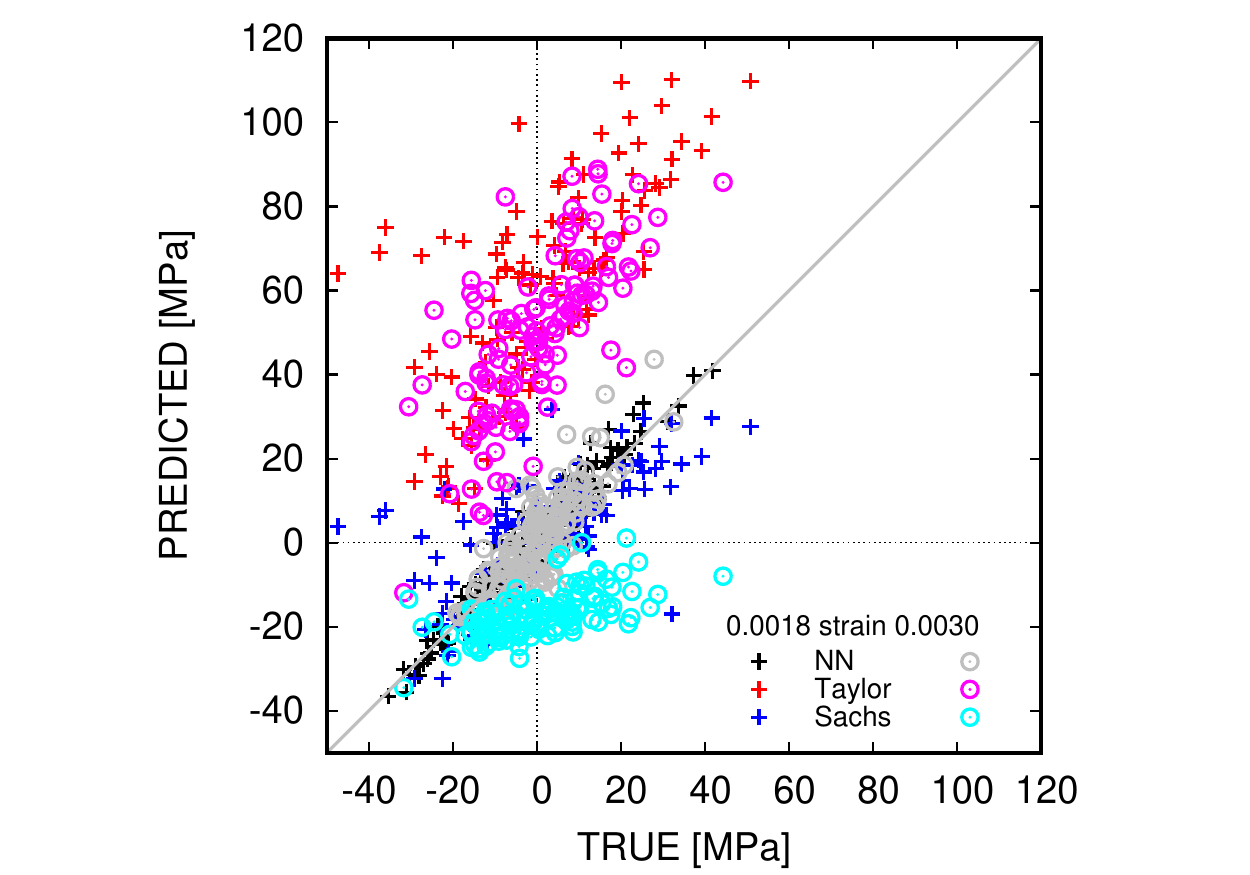}
\caption{Correlation of the ``true'' crystal plasticity deviations from the mean and neural network predictions as a function of strain, 0.0018 is at yield and 0.0030 is post-yield.
}
\label{fig:plasticity_correlation}
\end{figure}

\fref{fig:plasticity_correlation} compares the NN correlation with that of data produced with the Taylor and Sachs models at near yield total strain 0.0018 and a fully plastic total 0.0030.
The correlations are $R$=0.91, 0.76 for the NN, $R$=0.71, 0.69 for the Taylor model, and  $R$=0.68, 0.60 for the Sachs model.
Clearly, the NN has better or comparable correlations to the two classical models, and with lower bias.
Interestingly, the Taylor model behaves in a similar fashion to the Voigt model in overestimating the stress, but the Reuss analog, the Sachs model, has better accuracy albeit with increased bias.
The NN is also superior to the  average of the two semi-analytical models (not shown for clarity).
The cost of evaluating the NN model is tens of milliseconds of compute time, orders of magnitude lower than the tens of minutes it takes to solve the coupled highly nonlinear equations needed to evaluate the Taylor and Sachs models.
Also it should be noted that the Taylor and Sachs models, unlike the NN, have full, \apriori knowledge of slip behavior in each crystal through the constitutive models described in \sref{sec:realizations}.

\section{Conclusion} \label{sec:conclusion}

We have shown that data-driven homogenization via a hybrid neural network architecture is capable of predicting the variability in mechanical properties of polycrystalline materials by correlating the underlying crystalline structure with the stress response.
We demonstrated that a hybrid convolutional-recurrent neural network was able to predict the elastic response, the onset of plasticity, and the subsequent fully developed flow.
The predictions of the neural network trained on initial microstructures were in excellent agreement in the elastic regime but degraded om accuracy as plastic flow began to dominate the deformation and errors in the recursive component of the model accumulated.
This degradation in accuracy is common in other, more traditional modeling techniques, such as models based on the evolution determined by ordinary differential equations.

These results are promising and suggest that similar neural network architectures may be used on more complex materials and realistic microstructures.
It is clear that predictions of plastic flow are more difficult and would benefit from acquiring additional data.
However, gathering the requisite amount of data and training the networks is computationally expensive, and thus it is helpful to consider architectures that may allow for lower-dimensional representations of the grain structure.
Autoencoders may be used to construct lower dimensional representations (in a {\it latent} space or manifold) of grain structures that would lead to more efficient network training.
Using the results of a CNN to derive interpretable features of the grain structures directly would also enable physical insight into the primary drivers of material variability and enable simple reduced order models.

An important next step is to apply our methodology to real, experimentally observed microstructures.
The synthetic oligocrystals considered in this work have grains that are at the upper bound of those found in real additively manufactured materials and exhibit variability from a single mechanism.
Predicting the stress-strain response for realistic grain structures and general loading conditions would allow us to deploy these models as surrogates for more computational expensive constitutive models.
These models may prove useful in predicting material failure and connecting microstructural features to likelihood of failure.

An equally important direction for future work is to understand how the proposed neural network is representing the image information and  modeling the response.
Study of the {\it interpretability} of the predictions of the neural network, particularly the latent space constructed by image based component, with statistical, sensitivity and other means is a rapidly growing field \cite{ribeiroSG16,lipton16a,Zhang_2018_CVPR}.

\section*{Acknowledgments}
We relied on DREAM.3D (http://dream3d.bluequartz.net), Albany (https://github.com/gahansen/Albany), and Keras (https://keras.io/) to accomplish this work.
This work was supported by the LDRD program at Sandia National Laboratories, and its support is gratefully acknowledged.
Sandia National Laboratories is a multimission laboratory managed and operated by National Technology and Engineering Solutions of Sandia, LLC., a wholly owned subsidiary of Honeywell International, Inc., for the U.S. Department of Energy's National Nuclear Security Administration under contract DE-NA0003525.
The views expressed in the article do not necessarily represent the views of the U.S. Department of Energy or the United States Government.


\end{document}